\documentclass[final]{ws-ijmpc}
\usepackage{graphicx}
\usepackage[T1]{fontenc}
\begin{document}

\markboth{Henryk Fuk\'s, Babak Farzad and Yi Cao}
{A model of language inflection graphs}

\catchline{}{}{}{}{}

\title{A MODEL OF LANGUAGE INFLECTION GRAPHS}

\author{HENRYK FUK\'S}
\author{BABAK FARZAD}
\author{YI CAO}
\address{Department of Mathematics and Statistics\\
            Brock University,
         St. Catharines\\ Ontario, Canada  L2S 3A1\\
hfuks@brocku.ca, bfarzad@brocku.ca, caoyi163@gmail.com}

\maketitle

\begin{history}
\received{3 July 2013}
\revised{27 November 2013}
\accepted{29 November 2013}
\end{history}

\begin{abstract}
Inflection graphs are highly complex networks representing
relationships between inflectional forms of words in human
languages. For so-called synthetic languages, such as 
Latin or Polish, they have particularly interesting structure
due to abundance of inflectional forms. We construct the simplest
form of  inflection graphs, namely a bipartite graph in which one group of
vertices corresponds to dictionary headwords and the other group to inflected
forms encountered in a given text. We then study projection of this graph
on the set of headwords. The  projection decomposes into a large number of 
connected components, to be called word
groups.  Distribution of sizes of word group exhibits
some remarkable properties, resembling cluster distribution in a lattice
percolation near the critical point. We propose a simple model
which produces graphs of this type, reproducing the desired component distribution
and other topological features.

\end{abstract}

\ccode{PACS Nos.: 64.60.ah, 05.90.+m, 05.70.Jk, 02.50.-r, 64.60.aq}

\section{Introduction}
Human languages can be studied from many different perspectives. 
If we think of a foreign language, however, we typically think of \emph{words}
of that language, thus it is quite natural that vocabulary is one of the most
extensively studied features of languages. In recent years, the network paradigm
has been used to study vocabularies, and within this paradigm,  
words of the language are viewed as  vertices of a large and complex network or graph,
with edges representing
relationships between words. Many such models emphasizing different relationships between words
have been studied in the past decade,
including  networks of co-occurrences of words in sentences\cite{Ferrer2001},
thesaurus graphs\cite{MotterdLD02,KinouchiMLLR02,HolandaPKMR04}, WordNet database graphs\cite{citeulike:1179006},
and many others\cite{KeY08,CaldeiraLANM06,PomiM04,CanchoSK04,AntiqueiraNOC07}.

It is fair  to say that a lot of the aforementioned works concentrated on the English language, which
has a very characteristic property of being analytic, that is,  exhibiting  only a minimal inflection. 
In analytic languages grammatical relations and categories are handled mostly by the word order, and not by the inflection,
thus making them somewhat easier to learn. 

In contrast to this, synthetic languages such as Latin, Greek, Polish, or Russian make an extensive use 
of inflection, and one word in these languages can appear in great many forms, 
reflecting grammatical categories such as tense, mood, person, number, gender, case, etc.
Order of words is less important in synthetic languages.
While this is an excellent feature from the point of view of a poet,
it presents  algorithmic problems in text processing. Let us suppose, for example,
that we want to count the number of distinct words in a given work -- e.g., for the
purpose of comparing two works and deciding which one uses larger vocabulary.
  How do we do this in a language like Latin, where 
one dictionary headword can have as many as hundred different forms? To make things even more difficult,
in some cases, one inflectional form can correspond to more than one dictionary headword,
and one must deduce from the context which one to choose.

In Ref.~\refcite{paper38}, one of the authors considered this problem from a practical point of view,
and proposed a solution which exploits some features of the so-called inflection graph.
Here we will not dwell on this problem, referring an interested reader to Ref.~\refcite{paper38},
but we will instead discuss the inflection graph itself. We will first describe some of
its topological features, and then propose a model which reproduces these features.

\section{Inflection graphs}
The inflection graph for a given language can be constructed as follows. First we need to 
create a list of all words of the language, which, strictly speaking, is an impossible task,
as every such list is bound to be incomplete.
Nevertheless, one can easily obtain a reasonably adequate list of words using sufficiently large dictionary of the
language. The set of all dictionary headwords will be denoted by $H$. For each
headword, we generate a list of all possible inflected forms, and the list of all possible
inflected forms obtained this way will be denoted by $I$.  We then construct a bipartite graph 
$G=(H,I, E)$, where $E$ is the set of edges such that  the edge between $v\in H$ and $ u \in I$ 
exists if and only if $u$ is an inflected form of $v$.

Construction of the inflection graph is obviously possible only if one is able to produce
all inflected forms of a given word. For the Latin language, this can be achieved using WORDS,
a computerized dictionary of Latin created by William Whitaker\cite{words}. The resulting bipartite
graph has $1\, 028\, 972$ vertices and $1\, 077\, 806$ edges, and will be denoted by $G_{LA}$.

We were also able to construct inflection graph for Polish language, using lexical grammar
developed by the
 Group of Computer Linguistics of AGH University of Science and
Technology in Krak\'ow\cite{pisarek2009}. The corresponding graph, to be denoted by $G_{PL}$, has
 $1\, 872\, 140$ vertices and $802\, 911$ edges.

Normally, for most headwords in $H$, there are many corresponding
inflected forms in $I$, so an element of $H$ is typically connected to many (sometimes 100 or more)
elements of $I$. For example, the Latin word \textit{dicunt} (they say) and \textit{dixit} (he said) are both inflected forms
of the verb \textit{dico}, thus we will have a vertex in  $H$ corresponding to \textit{dico}
connected to vertices in $I$ corresponding to \textit{dicunt}  and \textit{dixit}.
However, the opposite can also be true: in some instances, a word
can be an inflected form of more than one headword, so that vertices of $I$ are sometimes
connected to more than one vertex of $H$. As an example, consider 
the word \textit{sublatus}, which could be a form of \textit{tollo} (lift, raise) or \textit{suffero} (bear, endure),
thus a vertex in $I$ corresponding to \textit{sublatus} will be connected to two vertices in $H$.

The inflection graphs are rather sparse, and they decompose into a large number of connected components of 
different sizes. From the practical point of view, the size of the component is not as important as the number of distinct
headwords in the component, which we will call \emph{headword groups}. The motivation for this can be explained as follows.

Suppose that one wants to perform a computerized count of the number of different words occurring in a given text.
Obviously, one wants to count two different inflection forms of a given word as one and the same
word, or, to put this differently, one wants to know how many distinct dictionary headwords appear in the text (in
various inflected forms). However, since in languages with a complex inflection system a given inflection form
can sometimes belong to two (or more) different dictionary headwords, it is impossible for a computer to
decide which one is used in a particular case. To make such a decision, one has to understand the sentence and
figure out from the context what is means. In English this problem is quite rare, but
still exists. For example, consider the word \emph{dove} -- this could be the singular form of the noun
\emph{dove} (a type of bird), or the simple past tense of the verb \emph{to dive}. Computer program upon encountering
\emph{dove}  in a text will not know whether to count it as occurence of the headword \emph{dove} or
\emph{to dive}. The simple solution to this problem is to say that \emph{dove} is an inflected form
of a headword from the set (headword group) \{\emph{dove, to dive}\}. This means that instead of counting how many
distinct headwords are present in the text, we can only count how many distinct  \emph{headword groups} 
are present.
We want to know, however, what are the sizes of the headword groups, as it is, in a sense, 
a measure of the difficulty of the disambiguation problem. A good way to analyze these sizes is
to look at their distribution. 

\begin{figure}
\centering
\includegraphics[width=2.4in]{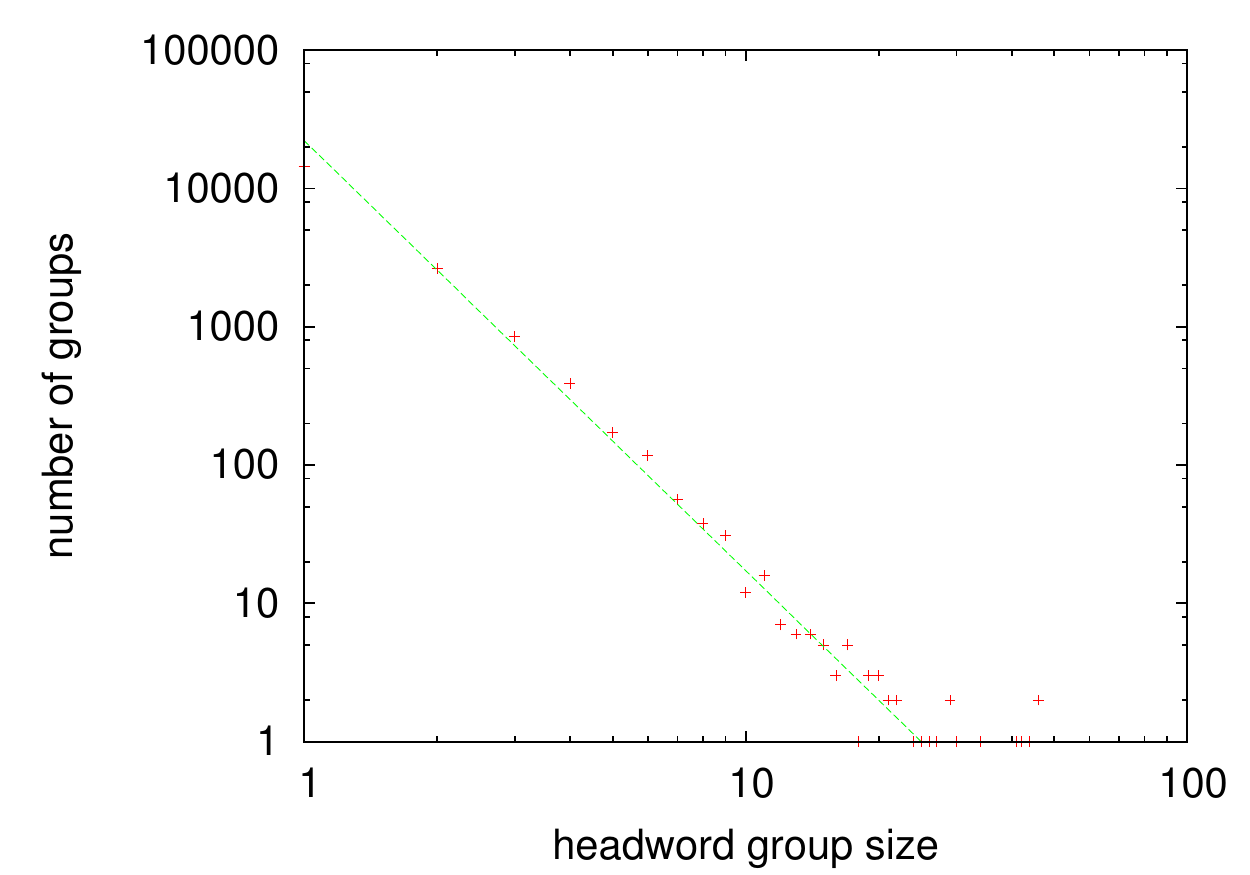} \includegraphics[width=2.4in]{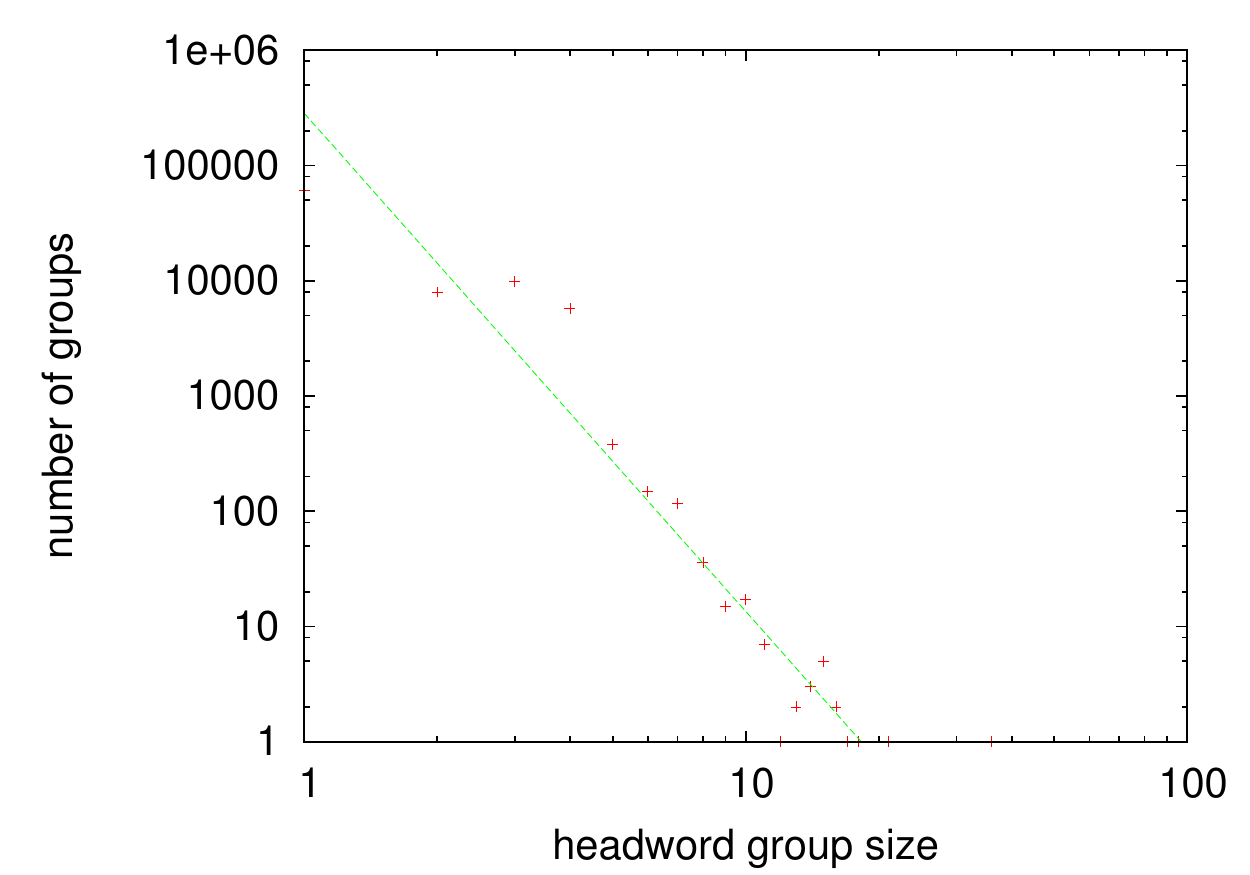}
\caption{Distribution of headword clusters for Latin (left) and Polish (right).
Slope of the fitted line is, correspondingly, $-3.1 \pm 0.3$ and  $-4.3 \pm 0.9$. The figure for Latin 
previously appeared in Ref. {\protect \refcite{paper38}}}\label{disthlatpol}
\end{figure}

The distribution of headword groups sizes in inflection graphs is quite striking, as can be seen in Figure~\ref{disthlatpol},
which shows the distributions for Latin and Polish. The graph for Latin and 
its analysis have been previously published in
Ref.~\refcite{paper38}, here we add the same graph for Polish language.

We fitted a straight line in log-log coordinates to data points for which the number of groups exceeds 20, in order to exclude
points with small count. The lines of the best fit are shown as dashed lines. 
There seems to be a power-law trend in both data, 
more strongly pronounced in the graph for the Latin language. In the remaining part of this paper we will attempt 
to shed some light on the origin of this phenomenon. 

The dashed lines of the best
fit shown in  Figure~\ref{disthlatpol} represent
the power law
\begin{equation}\label{powerlaw}
n_s \sim s^{-\tau},
\end{equation}
where $\tau \approx  -3.1 \pm 0.3$ for Latin and $\tau \approx -4.3 \pm 0.9$ for Polish. 
Errors given for $\tau$ signify that decreasing/increasing $\tau$ by the given amount
increases the reduced $\chi^2$ twice.
Anyone who is familiar with the percolation theory\cite{Stauffer92,Bollobas2006} can immediately recall that a very similar 
scaling law for cluster sizes holds for  the lattice  percolation at the critical point, where $\tau$
is known as the Fisher exponent\cite{Stauffer92}. This is also the case for the Erd\"os-R\'enyi model $G(n, p)$, that is, 
a graph constructed by connecting $n$ nodes randomly so that each edge is included in the graph with probability $p$
 independent from every other edge. 
\begin{figure}
\centering
\includegraphics[width=3.5in]{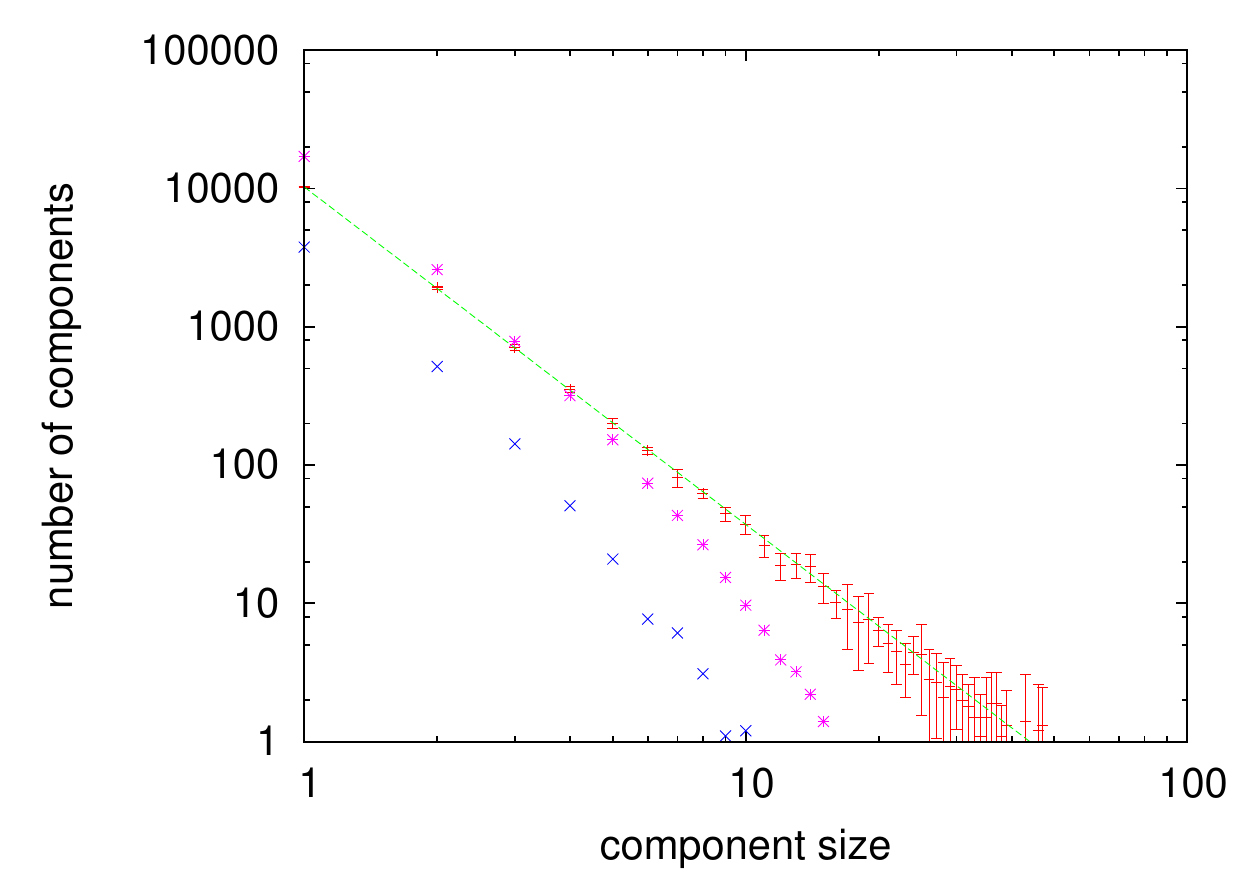} 
\caption{Distribution of component sizes for the random graph averaged over 10 realizations of the
graph above ($\times$), below ($\star$), and at the percolation threshold ($+$). The data point corresponding to
the giant component above the percolation  threshold is not shown. Slope of the fitted line is
 $-2.44 \pm 0.09$. Error bars correspond to standard deviations, and for clarity are shown only for the
data at the critical point.
}\label{percexper}
\end{figure} 
It is well known that at $np = 1$ and $n\to \infty$ the model undergoes a structural transition similar 
to percolation\cite{Bollobs2003}.
The distribution of component sizes follows the power law of eq. (\ref{powerlaw}), and the 
Fisher exponent is known\cite{Mori2001} to be $\tau=2.5$.
Figure~\ref{percexper} shows component size distributions obtained numerically for 
$G(n,p)$ with $n=28092$, that is, the same $n$  as the number of headwords in $G_{LA}$. Three values of $np$
were used, $np=0.5$ (below the percolation threshold), $np=2.0$ (above the percolation threshold)
and $np=1.0$ (at the percolation threshold). The power law in the form of eq. (\ref{powerlaw}) is evident
at the percolation threshold, yet it is clearly  not valid away from the threshold.
In spite of the fact that the number of vertices is relatively small and that
only 10 graphs were generated, the value of the exponent $\tau=2.44 \pm 0.09$ obtained
from fitting the straight line to data agrees, within error bounds, with the aforementioned
value of $\tau=2.5$.

Considering the case of $G(n,p)$, one could suspect that the inflection graphs have a structure somewhat resembling
Erd\"os-R\'enyi random graphs at the percolation threshold. We will, however, demonstrate that this is 
somewhat more complicated. To avoid repetitions, from now on we will be 
 using $G_{LA}$ as an example.

\section{Structure of the inflection graph for Latin}
In order to describe some important features of  $G_{LA}$,  we will consider its projection on $H$.
 Given a bipartite  graph $G = (H , I, E)$, define its $H$-projection
 as $G^\prime = (H, E^\prime )$, where $\{u,v\}$ is in $E^\prime$ if and only if $u$ and $v$ are both
connected to a common vertex in $I$. $H$-projection of $G_{LA}$ has 
 28092 vertices and 24064 edges. Only 13345 headwords have degree greater than zero in $G_{LA}^\prime$.
Note that for obvious reasons, distribution of component sizes of $G_{LA}^\prime$ is the same as the distribution of group sizes
in $G_{LA}$. Could it then be that $G_{LA}^\prime$ resembles Erd\"os-R\'enyi random graph?

\begin{figure}
\centering
\includegraphics[width=2.4in]{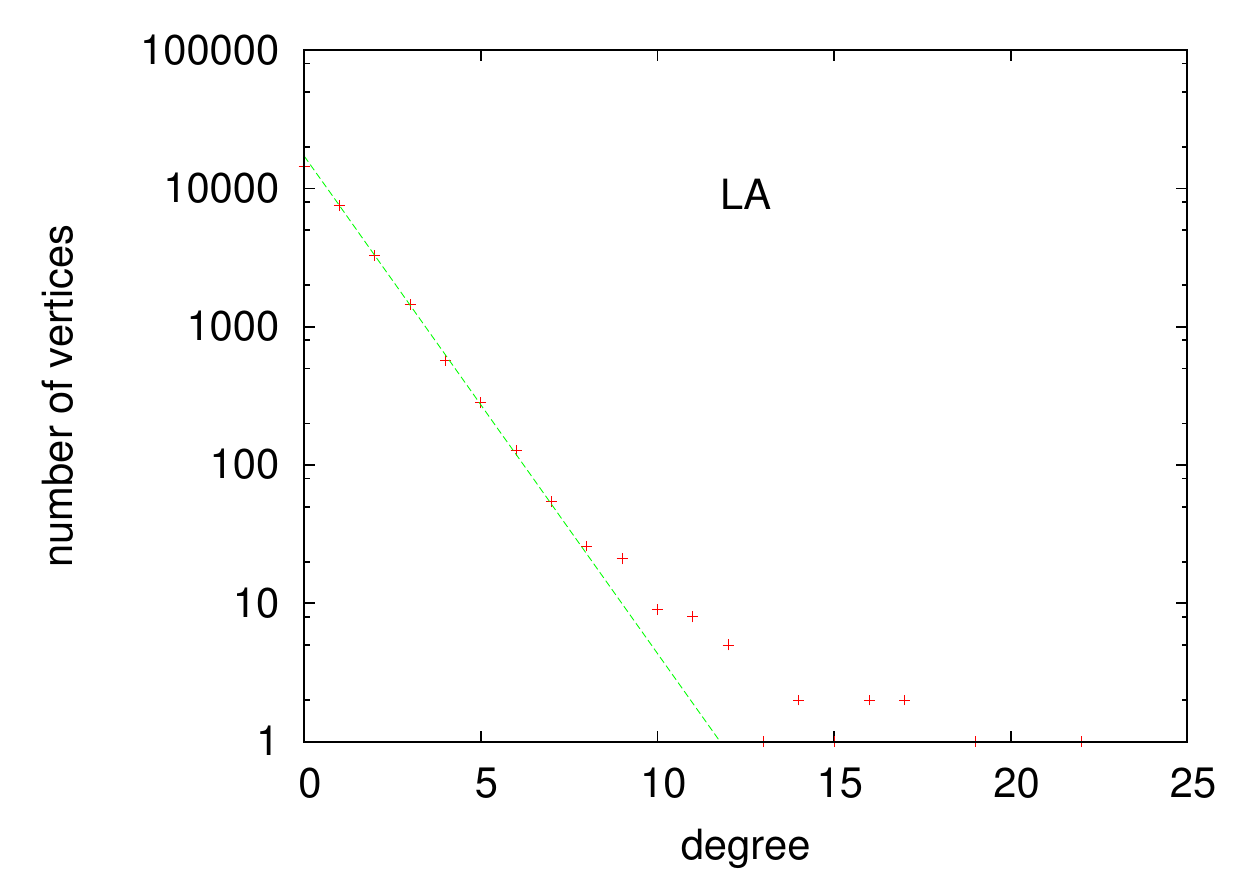} \includegraphics[width=2.4in]{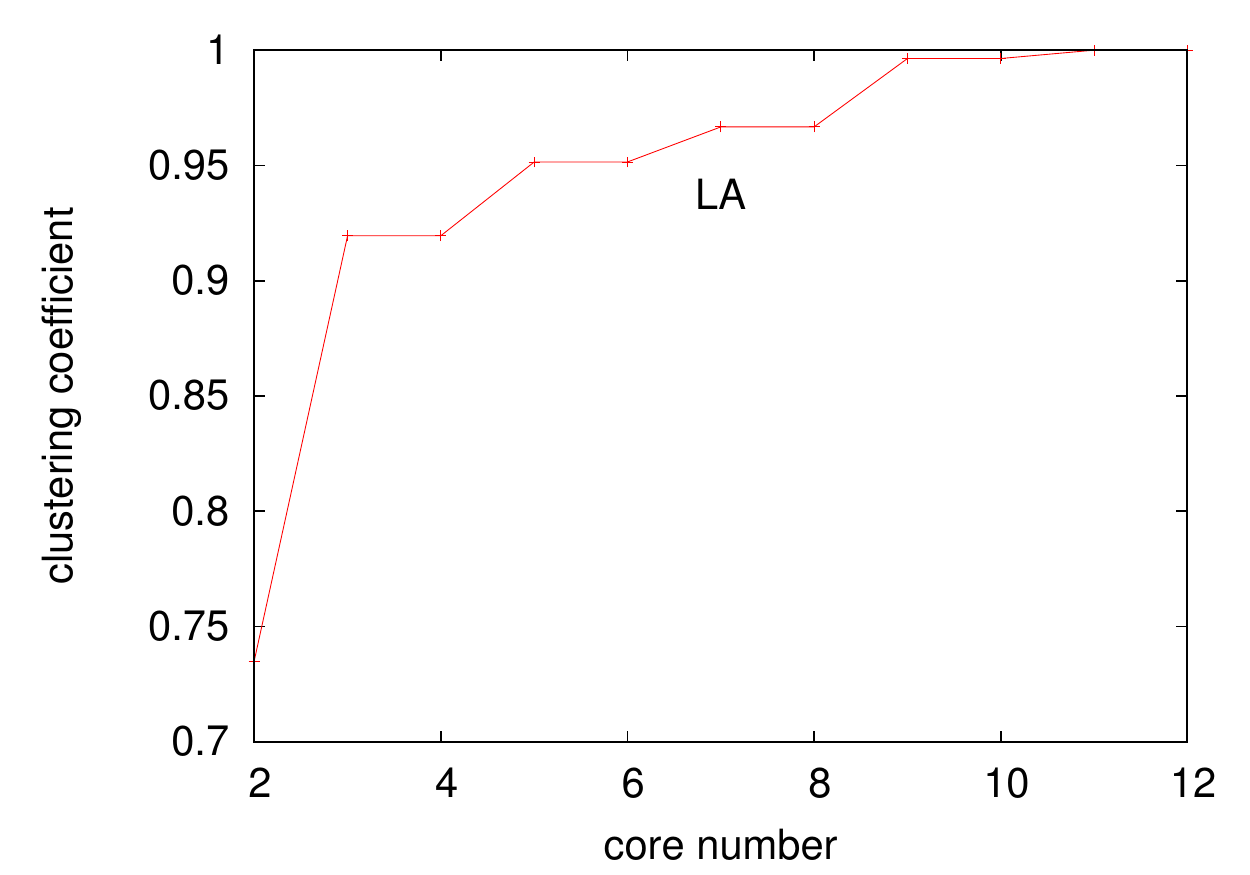}\\
\includegraphics[width=2.4in]{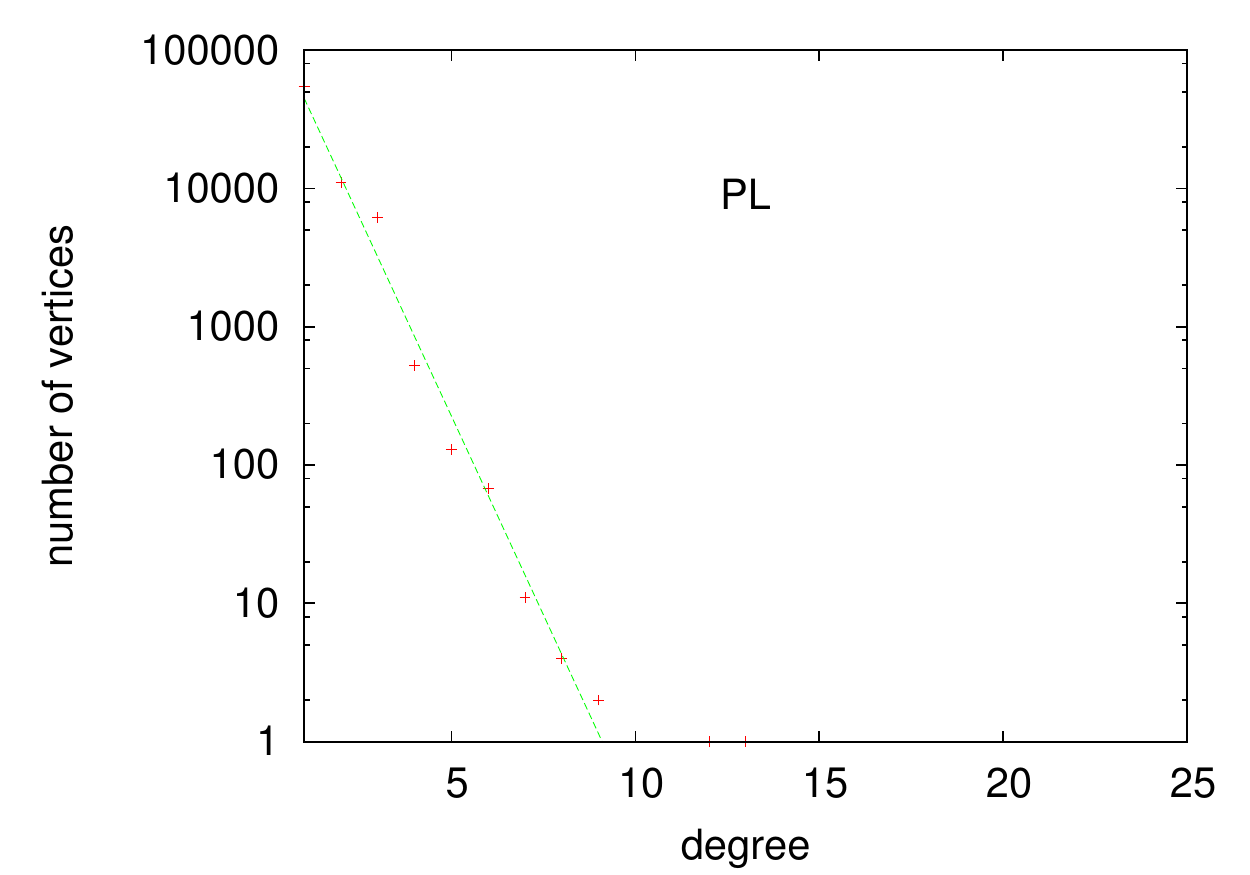} \includegraphics[width=2.4in]{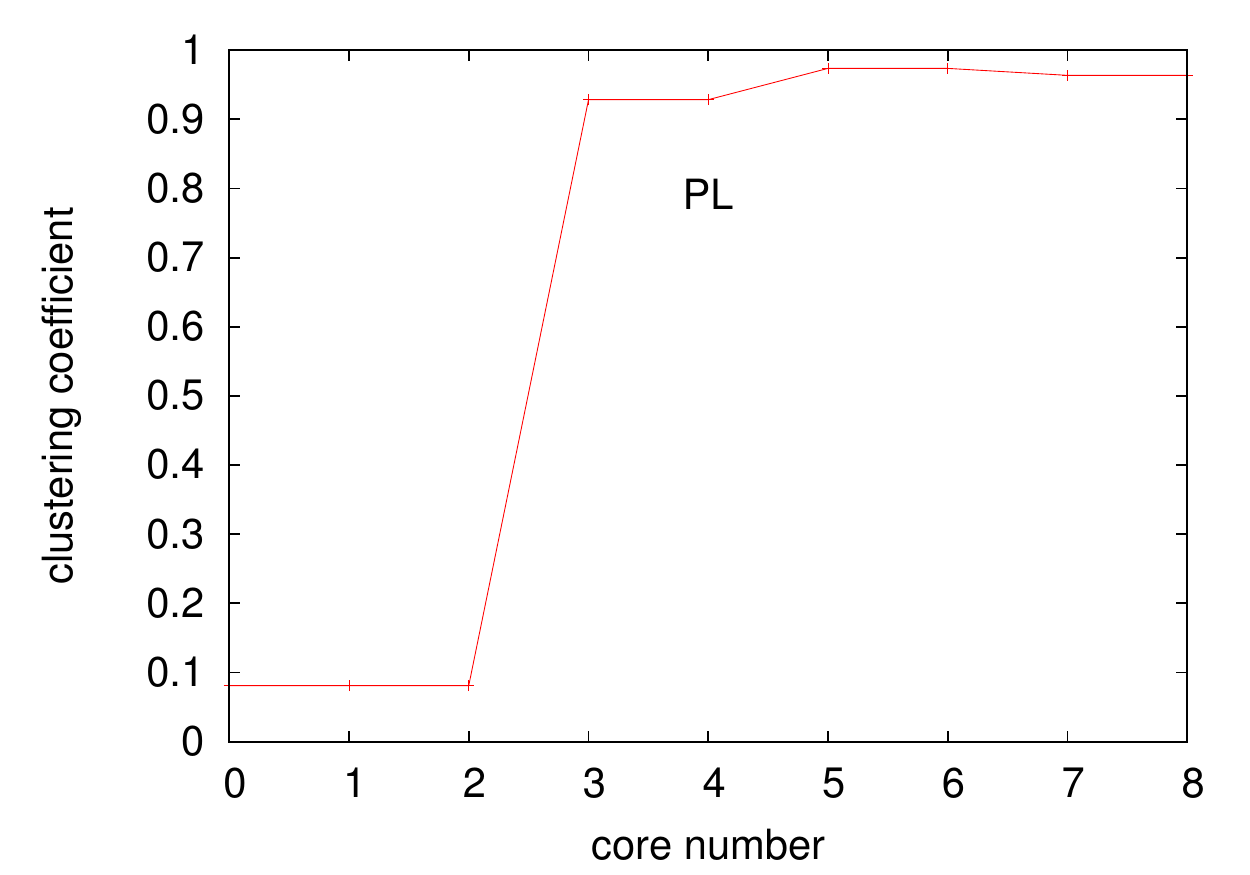}
\caption{Degree distribution (left) and clustering coefficients of cores (right) of $H$-projection of $G_{LA}$ (top)
and $G_{PL}$ (bottom).} \label{degdistproj}
\end{figure}

In order to answer this question, we will first consider the degree distribution of $G_{LA}^\prime$ shown in
Figure~\ref{degdistproj}. Unlike in the case of $G(n,p)$, the degree distribution
of $G_{LA}^\prime$ is clearly not Poissonian, and 
for small degree values it seems to follow exponential decay, shown in the figure as a straight line. 
The mean vertex degree is $1.8$. This already indicates that $G(n,p)$ cannot be a model
of $G_{LA}^\prime$ -- the mean vertex degree of $G(n,p)$ with a power law distribution
of components sizes must be equal to 1.0.

We can see the difference between $G(n,p)$ and $G_{LA}^\prime$ even better if we use the notion of \emph{core clustering spectrum},
introduced in Ref.~\refcite{paper37}. 
For a non-negative integer $k$, the $k$\emph{-core} of a graph is the maximal subgraph such that its vertices have degree
greater or equal to $k$. By the ``degree'' in this definition we mean the degree of the vertex in the subgraph.
 If $G$ is a given graph, we denote by $G_{\{k\}}$ the $k$-core of $G$. 
Now let $C(G)$ denote the clustering coefficient of $G$.
A set of pairs $(|G_{\{k\}}|,C(G_{\{k\}}))$, where $|G|$ denotes the number of vertices of $G$, will be called
core clustering spectrum of $G$. 
One can visualize the core clustering spectrum  by plotting points  $(|G_{\{k\}}|,C(G_{\{k\}}))$ on a plane, as
it has been done in Ref.~\refcite{paper37}. Here we will use  slightly different graphs in order to convey
a similar information, namely we will plot $C(G_{\{k\}})$ as a function of $k$. We will call it
the \emph{graph of clustering coefficients of cores}.  This has the advantage over
the plot of core spectrum in having the core number explicitly as one of the variables.
 The value of $k$ will range from $1$ to $k_{max}$, where $k_{max}$ is the largest $k$ for which $G_{\{k\}}$ is non-empty.

For some graphs, such as the Erd\"os-R\'enyi random graphs, most vertices belong to the same $k$-core,
as documented in Ref.~\refcite{Alvarez2005}. This means that the graph of clustering coefficients of cores for Erd\"os-R\'enyi random graphs 
is very narrow,  consisting of only a small number of points.
This is not the case for $G_{LA}^\prime$, as  Figure~\ref{degdistproj} attests.  $G_{LA}^\prime$ possesses highly clustered inner
core, feature absent in Erd\"os-R\'enyi model near the percolation threshold. 

Degree distribution of $G_{PL}^\prime$  and its graph of clustering coefficients of cores are quite similar to corresponding graphs
of $G_{LA}^\prime$, as shown in the bottom of Figure~\ref{degdistproj}. 
\section{Model}
In order to construct a model of inflection graphs which exhibits power law
scaling resembling Figure~\ref{disthlatpol}, as well as having the  degree distribution and clustering coefficients of cores
of  its $H$-projection resembling  Figure~\ref{degdistproj}, we need to make a couple of further 
remarks  regarding topological structure of inflection graphs, again using 
$G_{LA}$ as an example. It is useful to think of $G_{LA}$ as a collection of stars,
each centered at a headword and with arms connecting the 
headword to some inflected forms. These stars are not completely disjoint, however. Sometimes they 
share one or more vertices in $I$, and this occurs if a given headword shares some of its inflected forms
with another headword (or headwords).

Let $n$ be the number of
headwords, and $m$ be the number of inflected forms. Construction of the random graph serving as a model of $G_{LA}$ proceeds in two stages. 
In stage 1, we generate an assembly of stars, each centered at a headword and with arms connecting the 
headword to some inflected forms. In stage 2, we generate a number of random bridges between these stars. 
We now describe the two stages in detail.\\ \\
{\bf Algorithm for generating stars}
{\it
\begin{enumerate}
 \item Generate the set of vertices $H=\{H_1,H_2,\ldots, H_n\}$ corresponding to headwords, and another
set $I=\{I_1,I_2,\ldots, I_m\}$ corresponding to inflected forms.
\item For each $i\in \{1,2,\ldots n\}$, draw a random number $x_i$  from a distribution $f_h$ to be described below,
and connect vertex $H_i$ to vertices $I_{j+1}, I_{j+2}, \ldots,$ $I_{j+\lfloor |x_i| \rfloor}$,
where $j=0$ for $i=1$ and
$j=\sum_{p=1}^{i-1}{\lfloor |x_p| \rfloor}$ otherwise. 
If any vertex index in $I_{j+1}, I_{j+2}, \ldots,$ $I_{j+\lfloor |x_i| \rfloor}$ exceeds $m$, it is
replaced by its value modulo $m$. 

\item If any isolated vertices in $I$ still remain, connect each of them to a randomly selected 
vertex in $H$.  After this is done, relabel the set $I$ so that vertices connected to the
same headword are labeled with a block of consecutive integers. 
\end{enumerate}
}
The probability distribution function $f_h$ is a weighted sum of three normal distributions,
\begin{equation}
 f_h(x)=\sum_{i=1}^{3} w_i f_{\sigma_i, \mu_i}(x),
\end{equation}
where 
\begin{equation}
 f_{\sigma,\mu}(x)=\frac{1}{\sigma\sqrt{2 \pi }} e^{\frac{-(x-\mu)^2}{2 \sigma^2}}.
\end{equation}
We used values $(w_1,w_2,w_3)=(0.68, 0.28, 0.04)$, $(\mu_1,\mu_2,\mu_3) =(8,90,3)$
and $(\sigma_1,\sigma_2,\sigma_3)=(2,10,1)$. These were obtained by fitting the resulting degree
distribution to the degree distribution of the actual inflection graph, but their values are not
too critical, meaning that small changes in values of these parameters still produce graphs with
power-law distribution of headword group sizes.

Note that although the random number $x_i$ drawn from the distribution $f_h$ in step two may theoretically be zero, yet
the probability of such event is extremely small. In our program implementing the algorithm 
for generating stars, we  simply reject  $x_i=0$ outcome and draw another number if it happens. 

 The reason for taking $f_h$ to be the sum of three normal distributions is the structure of Latin vocabulary.
With respect to inflection, one can distinguish three main groups of words: (1) verbs (inflexion by conjugation), (2) nouns and adjectives
(inflection by declension) and (3) all other words. We should remark here that this shape of the distribution
is suitable for Latin, but for a different language, with a different grammatical structure,
it would have to be different -- in particular, the number of normal distributions in the sum would likely have to change.
Moreover, we used normal distribution for the sake of simplicity, and we do not claim
that this reflects the actual distribution of inflection forms very accurately, but it is close enough for our purposes.
 One should also note that $f_h$ may theoretically produce negative numbers (again, with very small probability), and this is why we take the absolute value
of $x_i$. We also round $x_i$ down to the nearest integer. One could use in place of the normal distribution
 some other distribution with 
strongly pronounced peak and producing
only positive numbers, such as, for example, the log-normal distribution. We found, however, that the detailed shape of the
distribution is not too crucial for our goal of reproducing the desired properties of the inflection graph, thus we 
kept the normal distribution for simplicity.

Once the assembly of stars is created, we add a number of bridges between the stars. The most crucial 
feature of these bridges comes from the fact that typically two headwords share not one, but
many inflected forms with another headword or headwords. This is because there exists a large number of pairs of closely related
Latin words, each having a separate entry in the dictionary. For example, the words \emph{dico} (say),
\emph{dictum} (utterance, remark) and \emph{dictus} (speech) are all closely related, thus they share
many inflected forms. After experimenting with many possible methods for generation of bridges, we 
came out with a simple algorithm, which basically adds a fixed number of edges at a time.

 Let $\lambda$ and $T$ be two positive integers, to be used as  parameters in our algorithm.\\\\
{\bf Algorithm for generating bridges}
{\it
\begin{enumerate}
\item Randomly select two headword vertices $H_a$ and $H_b$, where by $H_a$ we denoted the vertex with the
larger degree.   Vertex $H_a$ is already connected to $k$ inflected forms, let us
denote them by  $\{I_r, I_{r+1}, \ldots I_{r+k-1}\}.$ 
\item  Add $\lambda$ additional edges by 
connecting $H_b$ with vertices $\{I_r, I_{r+1}, \ldots I_{r+\lambda-1}\}$.
\item Repeat the above two steps $T$ times.
\end{enumerate}
}

Note that the second step is performed exactly as described even if  $\lambda>k$, but in this case some of the inflected forms 
with which we connect $H_b$ will not be inflected forms of $H_a$, but inflected forms of some other word(s). Also note that
$k$ is always greater than zero, because the algorithm for generating stars ensures that this is the case. This agrees with our
interpretation of the meaning of the ``inflected form''. We assume that every word has at least one inflected form -- if it is
an adverb, for example, its sole ``inflected form'' is identical to itself. This is consistent wit the treatment of other
parts of the speech. For instance, for nouns we count nominative singular among inflected forms, even though it is 
identical to the headword form.

Regarding the value of $\lambda$ and $T$, they must be selected as follows.
After completing the algorithm for generating stars, the number of edges in the graph is only slightly larger than $|I|$
(recall that in step 2 we are replacing indices exceeding $m$ with their values modulo $m$, but this happens only rarely for
a few values of  $i$ close to $n$).  Of course it could theoretically happen that the number of vertices
will be larger than the desired number of edges (we want to have the same number of edges in the model graph
as in the inflection graph being modeled). With the choice of parameters which we have made, 
the probability of such event is 
so exceedingly small, that for all practical purposes we can simply ignore such eventuality. Nevertheless, if it indeed happened,
one would have to discard the result and run the algorithm for generating stars again.

Having less than the desired number of edges, we must ensure that the product  $\lambda T$ is equal to the
number of remaining edges which we want to produce.  This means that only one of those two parameters can be freely chosen.
By experimenting with
different values of $\lambda$ in the range from $1$ to $15$, we found that $\lambda=10$ produces the most clearly pronounced
power-law distribution of headword sizes in the resulting graph. The typical corresponding value of $T$ in this case is $T=7692$.
We say ``typical'' because, as explained earlier, the exact number of vertices in the graph obtained after
applying the algorithm for generating stars will slightly fluctuate for different realizations of the graph, thus
the number of ``missing vertices'', and consequently the value of $T$, will slightly fluctuate too.
The shape of the headword group size distribution graph, 
however, is only weakly affected by changes of $\lambda$ and $T$ as long as their product remains equal to
the number of ``missing vertices''
 and providing that  $\lambda>1$. 
For example,  if instead of  $\lambda=10$ and $T=7692$ we use $\lambda=5$ and $T=15384$,  there is almost no perceptible difference 
in the shape of the graph. 

\begin{figure}
\centering
\includegraphics[width=3.5in]{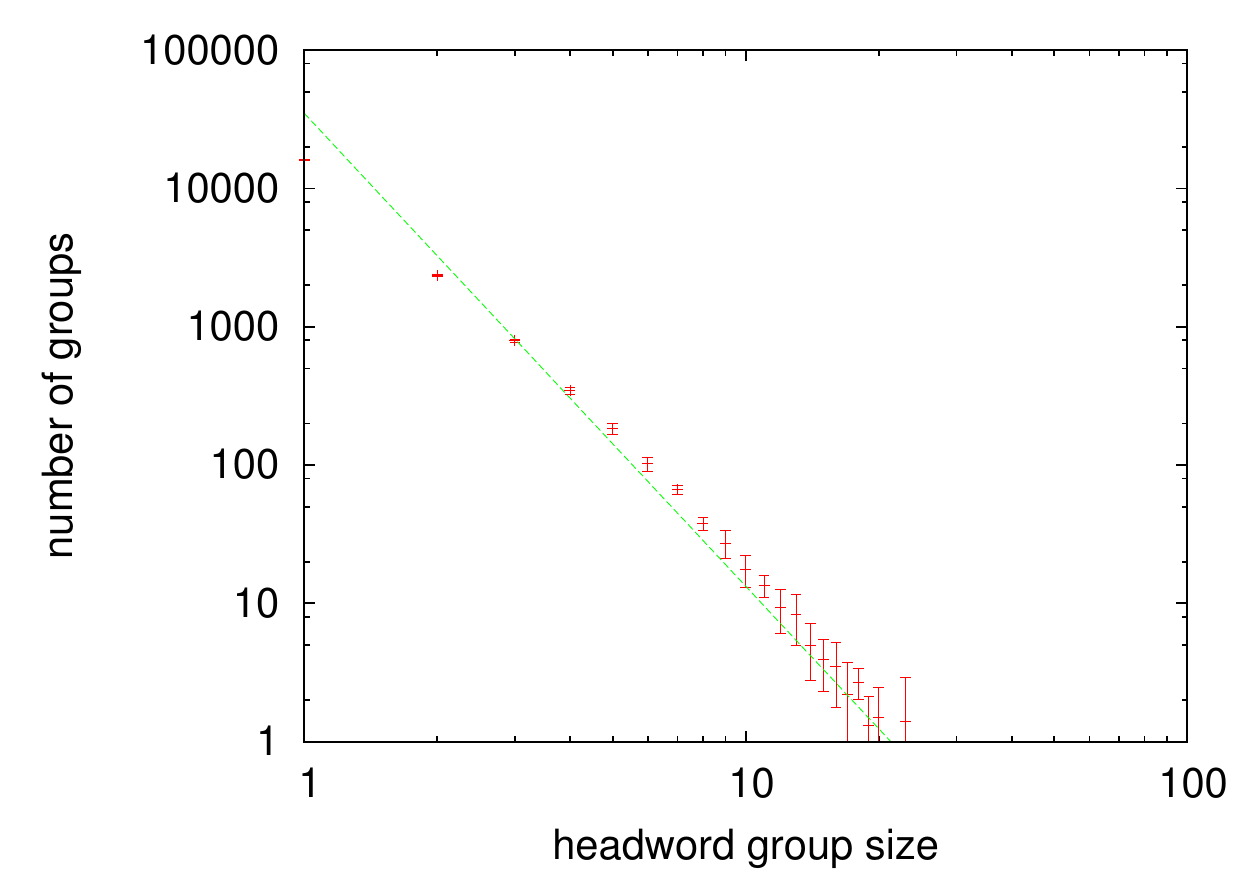} 
\caption{Distribution of headword group sizes for the model graph $G_{MOD}$ averaged
over 10 realizations of the graph.
Slope of the fitted line is $-3.4 \pm 0.5$.} \label{model-groups}
\end{figure}

We generated random graph
following the above algorithm using $|H|=28092$ and $|I|=1000880$, that is, the same number of vertices as in the actual inflection graph.
This graph will be called $G_{MOD}$.
Its distribution  of headword group sizes is shown in Figure~\ref{model-groups}. Agreement with
the actual distribution shown in
Figure~\ref{disthlatpol} is indeed very good. Even the slope of the fitted line agrees (within the error bound) with the
exponent observed in $G_{LA}$, as these are respectively $-3.4 \pm 0.5$ and $-3.1 \pm 0.3$.

\begin{figure}
\centering
\includegraphics[width=2.4in]{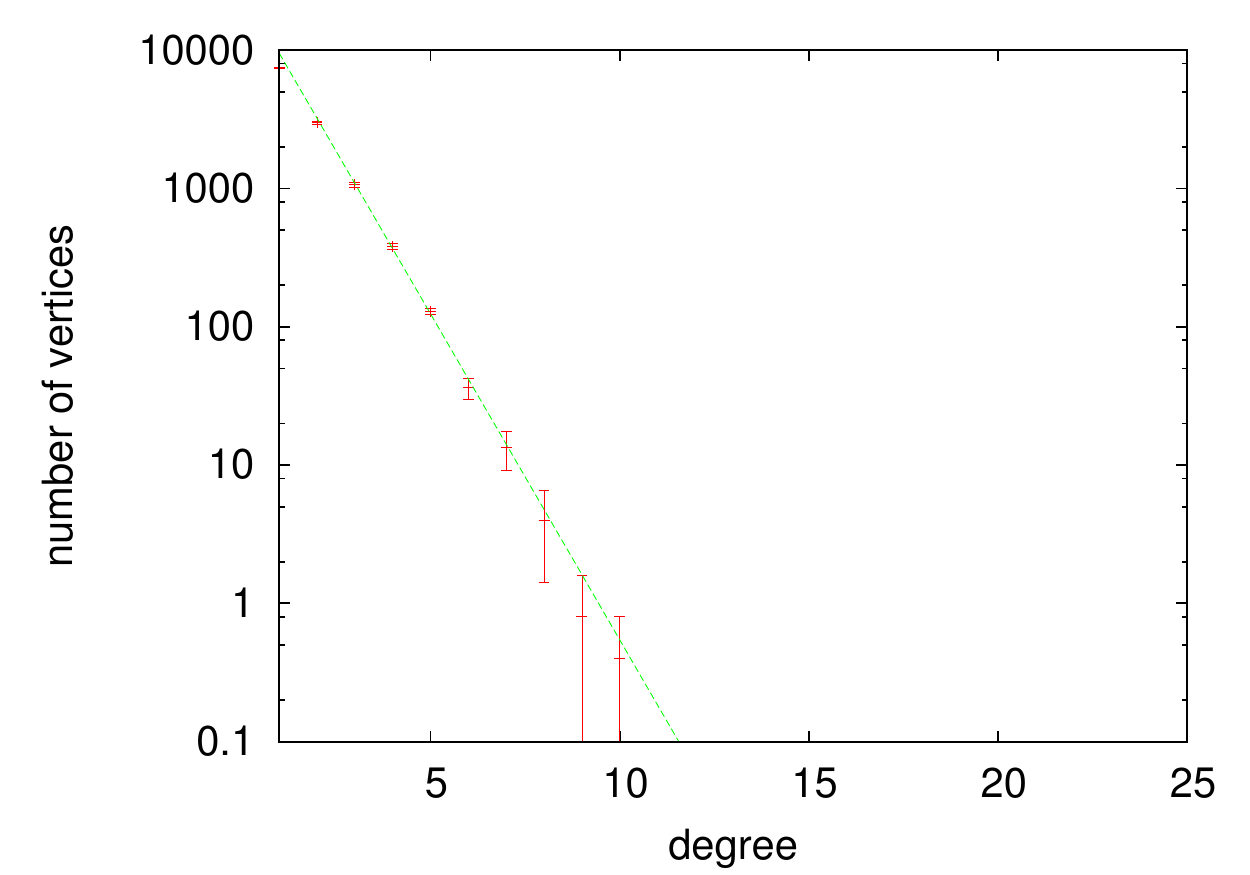} \includegraphics[width=2.4in]{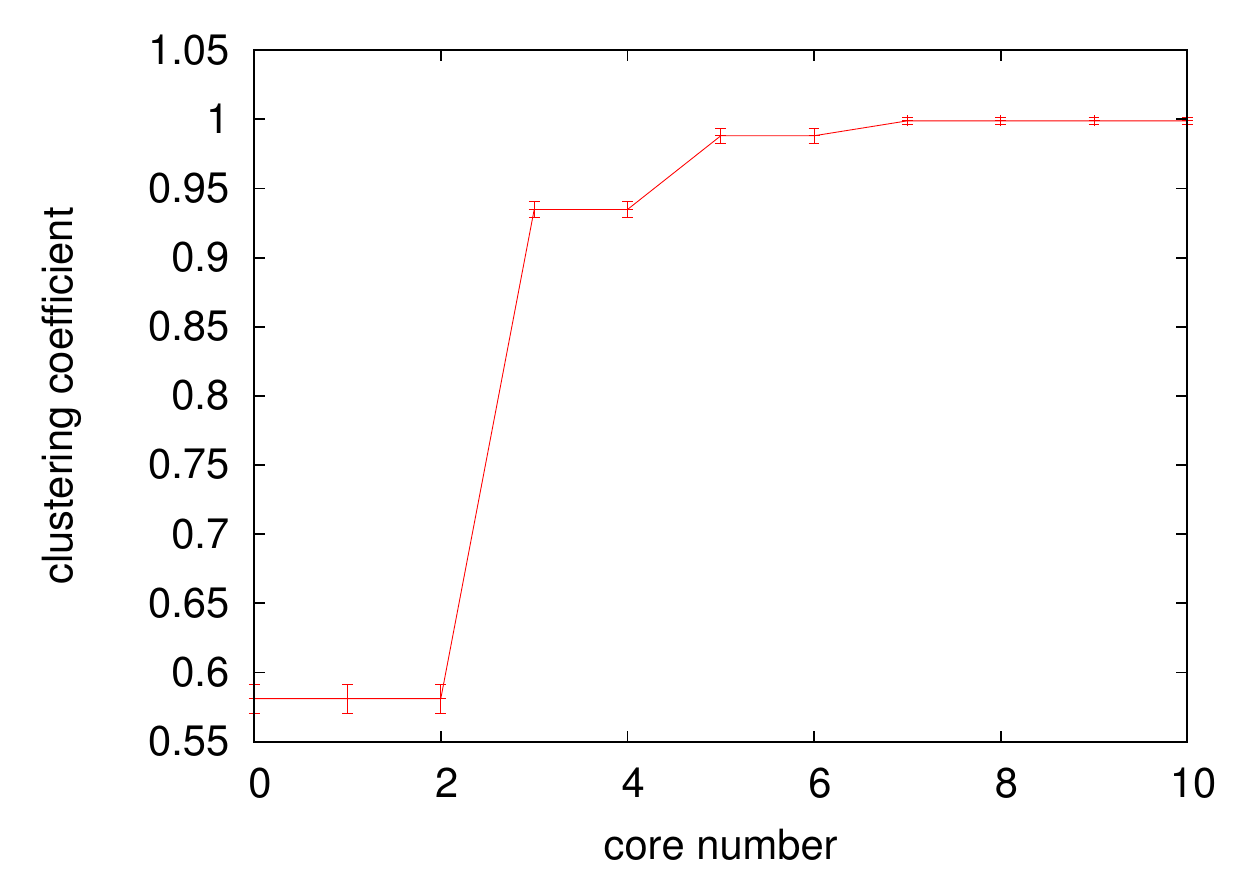} 
\caption{Degree distribution (left) and clustering coefficients of cores (right) of $H$-projection of the model graph,
averaged over 10 realizations of the graph. Error bars correspond to
standard deviation.} \label{degdistmod5}
\end{figure}

The model also performs well when one considers $H$-projection of $G_{MOD}$. Figure~\ref{degdistmod5}
shows both the degree distribution and the graph of clustering coefficients of cores of $G_{MOD}^\prime$.
Comparing these graphs with Figure~\ref{degdistproj}, we can observe good qualitative agreement. 
Degree distribution of $G_{MOD}^\prime$ is very similar to degree distribution of
$G_{LA}^\prime$, except that $G_{MOD}^\prime$ misses a small number of high-degree vertices, present
in $G_{LA}^\prime$. Clustering coefficients of cores of both graphs exhibit very similar behavior, that is, the clustering sharply increases
with increasing core number, and reaches value close to 1 for the inner core, indicating the presence of 
 cliques in high (innermost) cores.
\section{Conclusions}
We have discussed selected topological properties of inflection graphs and proposed 
a random graph model which exhibits the desired properties. In particular, our model
possesses nearly identical distribution of headword group sizes, and its $H$-projection
exhibits degree distribution and clustering coefficients of cores qualitatively similar
to analogous properties of the original inflection graph for the Latin and Polish languages.

A number of unresolved questions remain. First of all, it would be helpful to 
formally prove that the distribution of headword group sizes
in our model follows a power law, as well as to prove that the degree distribution
of the $H$-projection decreases exponentially with degree. We feel that further
simplification of the model may be needed in order to achieve this goal.

A separate question is the meaning and implications of the observed features of inflection graphs
in the linguistic context. It seems plausible, for example,  that the structure of the inflection graphs
is in some sense optimal. If the number of ``bridges'', that is, connections between headword stars
was much higher, the whole inflection graph would be connected, and the disambiguation of
headwords based on inflected forms would be difficult. On the other hand, if there were no bridges between headword
stars at all, then a much larger number of inflected forms would be needed. One can therefore speculate
that the actual inflection graph represents some sort of compromise between these two
extremes. In order to substantiate this claim one would need to construct a dynamical
process producing many possible forms of inflection graphs, and then show that the attractor of this process
is the actual inflection graph, just like in the case of self-organized criticality. 

It is also possible to draw some further analogy between the percolation process and inflection graphs.
One can think of percolation as a process in which one starts with a graph with $n$ vertices and no edges, 
and then adds random edges one by one. The graph will then undergo a percolation transition, and
the power-law distribution of component sizes will be observed at the transition point. Below and
above the percolation point, no power law will be observed. In order to mimic this process, we took
the graph  $G_{LA}$ and started adding random edges to it\cite{YiCao2011}. As expected, this 
destroyed the power-law distribution of components sizes of $G^{\prime}_{LA}$, although, obviously, 
it is very difficult to pinpoint how many edges \emph{exactly} are needed to destroy the power law -- 
the power law is not exact in the first place. The same phenomenon can be observed when one adds 
random edges to $G_{MOD}$. One can thus say that inflection graphs as well as the model graph
are somewhat ``frozen''  at the threshold, or slightly below the threshold, of some percolation process. 
As intriguing as it is, this  statement has to be taken very cautiously, because in the actual inflection 
graph edges cannot be added or removed -- the graph is a fixed feature of the language. 
We plan to probe this issue further  in the near future.

\footnotesize
\vskip 1cm
\noindent{\bf Acknowledgments}\\
H.F. and B.F. wish to acknowledge partial financial support from the Natural
Sciences and Engineering Research Council of Canada (NSERC) in the
form of Discovery Grant. This work was made possible by the facilities of the Shared
Hierarchical Academic Research Computing Network (SHARCNET:www.sharcnet.ca) and
Compute/Calcul Canada.  We also thank Dr. P. Pisarek for supplying the data needed for
construction of the  Polish inflection graph. 


\providecommand{\href}[2]{#2}\begingroup\raggedright\endgroup

\end{document}